\documentclass{aa}
\usepackage{graphicx}
\begin{document}
%
%
%
\newcommand{\ddp}[2]{\frac{\partial #1}{\partial #2}}
\newcommand{\ddps}[2]{\frac{\partial^2 #1}{\partial #2 ^2}}

\thesaurus{02.04.01, 08.14.1, 08.16.6, 02.07.01}
\title{Fast rotation of strange stars}
\author{E. Gourgoulhon\inst{1}  
 \and 
P. Haensel\inst{1,2,3} 
 \and 
R. Livine\inst{3} 
 \and 
E. Paluch\inst{3}
 \and 
S. Bonazzola\inst{1}
 \and 
J.-A. Marck\inst{1}
}
\institute{D\'epartement d'Astrophysique Relativiste et de Cosmologie 
-- UPR 176 du CNRS, Observatoire de Paris, F-92195 Meudon Cedex, France
\and
N. Copernicus Astronomical Center, Polish
           Academy of Sciences, Bartycka 18, PL-00-716 Warszawa, Poland
\and
Ecole Normale Sup{\'e}rieure de Lyon, 46, all{\'e}e d'Italie, 
69364 Lyon, France\\
{\em Eric.Gourgoulhon@obspm.fr, haensel@camk.edu.pl}}
\offprints{P. Haensel}
\date{received/accepted}
\maketitle
%
\begin{abstract}
Exact models of uniformly rotating strange stars, built  of self
bound quark matter, are calculated within the framework of
general relativity. This is made possible thanks to a new numerical
technique capable to handle the strong density discontinuity 
at the surface of these stars. Numerical
calculations are done for a simple MIT bag model equation of state
of strange quark matter. 
Evolutionary sequences of models of rotating strange stars at 
constant baryon mass are calculated. 
Maximally rotating configurations of strange stars are determined, 
assuming that the rotation frequency is  limited
by the mass shedding and the secular instability with respect to
axisymmetric perturbarions. 
Exact formulae which give the dependence of the maximum rotation
frequency, and of the maximum mass and corresponding radius of rotating 
configurations,  on the value of the bag constant,  are obtained. 
The values of $T/W$ for rapidly rotating massive strange stars 
are  significantly higher than those for ordinary neutron 
stars. This might indicate particular susceptibility of rapidly rotating 
strange stars to triaxial instabilities.

\end{abstract}
%

\section{Introduction}
A deconfined, beta stable  quark matter is composed of the u, d,
and s quarks,  and in contrast to ``ordinary'' baryon matter has
the strangeness per unit baryon number $\simeq -1$ (therefore
the name `strange quark matter').  An 
intriguing possibility that such {\it strange matter} could be
the absolute ground state of matter at zero pressure and
temparature was pointed out in an influential paper of Witten
(1984) [such a possibility has been contemplated
before  by Bodmer (1971)]. This 
possibility is not excluded by
what we know from laboratory nuclear physics. If the hypothesis
of strange matter is true, then some of neutron stars could be
actually strange stars, built entirely of strange matter
(Haensel et al. 1986, Alcock et al. 1986; arguments against the
existence of strange stars have been presented by Alpar (1987), 
 and  Caldwell \& Friedman (1991) ). 

Detailed models of nonrotating strange stars were constructed in
(Haensel et al. 1986,  Alcock et al. 1986). 
 Calculations concerning physics and astrophysics of strange
stars, published up to 1991,  were reviewed in (Madsen \& Haensel
1991). References to more recent  work on strange stars can be
found in   review articles by Weber et al. (1995) and Madsen (1999), 
and in the monograph of Glendenning (1997). 
Most recently, strange stars have been invoked in models of 
X-ray bursters (Bombaci 1997, Cheng et al. 1998) and 
of gamma-ray bursters (Dai \& Lu 1998). 

Rapid rotation of strange stars has become a topic of 
interest in 1989, after a sensational (but subsequently
withdrawn) claim of detection of a $0.5$~ms pulsar in SN~1987A. 
 In several papers, strange star has been advanced as
 an appropriate   model for a half-millisecond pulsar
(Glendenning 1989a,b, Frieman \& Olinto 1989). 
However, several authors pointed out
that the experimental constraints on the strange matter model
imply that strange stars could not rotate so fast (Haensel and
Zdunik 1989, Lattimer et al. 1990, Zdunik \& Haensel 1990, 
Prakash et al. 1990). 
Subsequent  studies of rapid rotation of strange stars focused on
the effect of rotation on the structure of the normal crust
(Glendenning \& Weber 1992). 

Recently, it has been shown that fast rotation of hot, young 
neutron stars is severely limited  by the emission of gravitational 
radiation, due to 
r-mode instability [for a review, see e.g. Stergioulas (1998)]. 
However, in contrast to newly born neutron stars, 
hot strange stars are not subject to the r-mode instability (Madsen 1998). 
Therefore, strange stars formed in collapse of rotating stellar 
cores can rotate very fast. 
 This gives additional motivation for studying rapid rotation 
of strange stars.  

It should be stressed, that nearly all calculations of uniformly
rotating models of strange stars were done within the slow
rotation approximation 
(see, e.g., Colpi \& Miller 1992, Glendenning \& Weber 1992). 
Actually, in the calculations of
Glendenning \& Weber (1992) the slow rotation scheme, pioneered by Hartle 
(1967, 1973) and
Hartle \& Thorne (1968), was supplemented by a ``self-consistency
condition'' at the Keplerian frequency (Weber \& Glendenning
1991, 1992). However, the slow rotation approximation is
not valid near the shedding (Keplerian) limit. As discussed by 
Salgado et al. (1994a, b),  the
improved, ``self-consistent'' version of this approximation,
overestimates by more than 10\% the maximum rotation frequency of
neutron star models. 

So far, the only exact\footnote{In this article, the term ``exact'' 
is relative to the treatment of rotation and is used
to distinguish from the slow rotation approximation}
calculations of the rapidly rotating
models of strange stars were done by Friedman (quoted  in
Glendenning 1989a,b) and Lattimer et al. (1990). 
However, the precision of  their numerical results,
calculated for one specific equation of state of strange matter,
was  not checked using an independent exact calculation.  Also,
the fact that their 2-D code was based on the Butterworth \&
Ipser (1976) scheme, rises some suspicion concerning its
precision when applied to the equation of state of strange
matter, characterized by huge density discontinuity at the
stellar surface. Indeed Butterworth \& Ipser (1976) considered
the case of a density discontinuity at the stellar surface -- they
computed configurations of rotating homogeneous bodies, but as we 
discuss below (Sec.~\ref{s:compar}), 
they treated properly the discontinuity only in the radial
direction, which is not sufficient for objects which deviate substancially
from the spherical symmetry.

In the present paper we perform exact calculations of rapidly
rotating strange stars, composed entirely of
strange quark matter. Our calculations are based on a multi-domain 
spectral method recently developed by Bonazzola et al. (1998b). 
Such a multi-domain technique enables us to treat exactly the density
discontinuity at the surface of strange stars. 
The calculations are performed for a
family of equations of state of strange quark matter. On
one hand, precision of our calculation is checked using internal
error estimators. On the other hand, the validity of scaling
relations for the rotating configurations, 
 displayed by our numerical results, yields an
additional test of precision of our calculations. Finally, we
address the problem of triaxial instabilities of rapidly rotating
strange stars, and point out differences with respect to normal
neutron stars. 

In Sect.~\ref{s:EOS} we discuss the equation of state of strange matter,
based on the MIT bag model of quark matter. 
The set of equations to be solved and the numerical procedure is presented
in Sect.~\ref{s:formul_code}. 
Stationary, uniformly rotating configuration of strange stars are studied
in Sect.~\ref{s:result}, where we also derive exact scaling relations for the
parameters of the rotating strange star models. Parameters of
the maximally rotating configurations, 
 stable with respect to the axially-symmetric perturbations,
are calculated in Sect.~\ref{s:rotmax},
and an exact formula for the maximum rotation frequency of
rotating strange stars is derived.  
  We derive also exact 
formulae for  mass and radius  of the maximum mass
configuration of rotating strange stars. In Sect.~\ref{s:bar_instab}, we
study  the problem of triaxial instabilities of rapidly rotating
strange stars, and show, that rapidly rotating strange stars might be
more susceptible to these instabilities than ordinary neutron stars.
 Finally, Sect.~\ref{s:concl} contains a discussion of our
results and the conclusion.
%
\section{Equation of state and static models of bare strange stars} 
\label{s:EOS}
Equation of state of strange matter will be  based on the MIT
bag model.  Baryon number density of strange matter is 
$n={1\over 3}(n_{\rm u} + n_{\rm d} + n_{\rm s})$, where 
$n_{\rm u}$ is the number density of the u-quarks etc.  
In what follows, we will use the simplest 
model of self-bound  strange quark matter, 
ignoring  the strange quark mass and neglecting the quark interactions  
except for the confinement effects described by the bag constant. 
We consider thus  massless, noninteracting  
u, d, s quarks, confined to the bag volume. In  the case of a 
bare strange star,  the boundary of the  bag coincides  
with stellar surface. 
 
Using the model described above, one can 
express the energy density, $\epsilon$, and the pressure, $P$, of
strange quark matter,  as functions of the baryon number density,
$n$, in the following form
\begin{eqnarray}
\epsilon &=& a n^{4\over 3} + B~,\nonumber\\
 P &=& {1\over 3}a n^{4\over 3} - B~,
\label{EOS_sm}
\end{eqnarray}
where $B$ is the MIT bag constant, and the parameter $a$ is
given by 
\begin{equation}
a = {9\over 4}\pi^{2\over 3}\hbar c = 952.371~{\rm MeV~fm}  \ .
\label{def_a}
\end{equation}
The bag constant $B$ describes the difference in the energy density
of the true (real) vacuum and that of the QCD vacuum. This
parameter plays a crucial role in the MIT bag model, being
actually responsible for the quark confinement. 
Let us mention, that
in the case of a bare strange star  of $\sim {\rm M_\odot}$ 
 we are dealing with a bag of a radius of $\sim 10~{\rm km}$, 
containing $\sim 10^{57}$ quarks.

\begin{figure}
\centerline{\includegraphics[width=8.5cm]{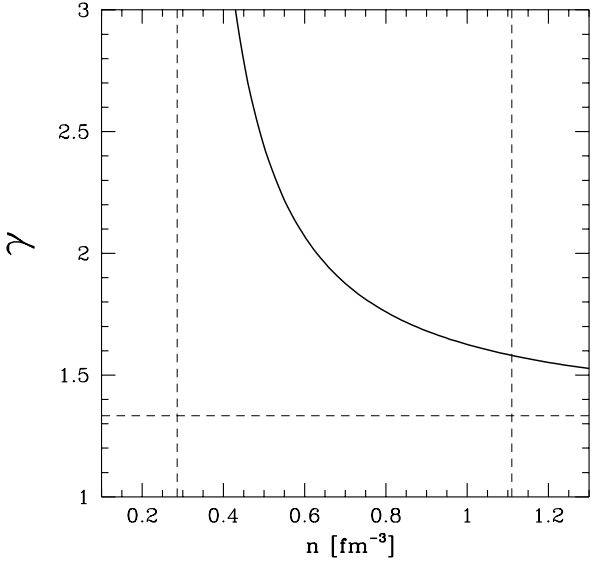}}
\caption[]{\label{f:gamma}
Adiabatic index of strange quark matter, $\gamma$, versus 
baryon density, $n$. Two vertical dashed lines correspond 
to strange star surface (left line), and maximum density 
within static strange star models (right line).
Horizontal dashed line corresponds to $\gamma=4/3$, characteristic 
of free ultrarelativistic Fermi gas.  
}
\end{figure}
The constant $B$ represents a natural unit for both the energy
density and the pressure. Defining dimensionless energy density and
pressure of strange matter, $\widetilde\epsilon \equiv
\epsilon/B$, $\widetilde P = P/B$, we can write down a
dimensionless form of the EOS of strange matter,
\begin{equation}
\widetilde P = {1\over 3}(\widetilde\epsilon - 4)~.
\label{EOS_tilde}
\end{equation}
The above dimensionless form of the EOS of strange matter constitutes 
a basis for the derivation of the {\it scaling laws} relating the 
families of static models of strange stars, 
calculated for different values of $B$  
(Witten 1984, Haensel et al. 1986). 
For the (gravitational) mass, baryon mass, and radius of static 
configuration with maximum allowable mass
we get 
(assuming  $G=6.6726~10^{-11} {\rm\ m^3 \, kg^{-1} s^{-2}}$, 
${\rm M}_\odot= 1.989~10^{30} {\rm\ kg}$
and $c=2.99792458~10^{8} {\rm\ m\, s^{-1}}$):
\begin{eqnarray}
M_{\rm max}^{\rm stat}&=&{1.9638\over \sqrt{B_{60}}}~{\rm M}_\odot~,
\nonumber\\
M_{\rm B,max}^{\rm stat}&=&{2.6252\over \sqrt{B_{60}}}~{\rm M}_\odot~,
\nonumber\\
R_{M_{\rm max}}^{\rm stat}&=&{10.712\over \sqrt{B_{60}}}~{\rm km} \,
\label{MR.stat}
\end{eqnarray}
where $B_{60}\equiv B/(60~{\rm MeV~fm^{-3}})$. The baryon mass is 
defined as $M_{\rm B}\equiv 1.66~10^{-27}~{\rm kg}\cdot N_{\rm B}$, where 
$N_{\rm B}$ is the total baryon number of stellar configuration.  


Consider the case of strange matter at zero pressure (such a
situation corresponds to the surface of a ``bare strange
star''). For massless, noninteracting  quarks, 
we obtain the following
set of parameters, characterizing the properties of strange
matter at zero pressure,
\begin{eqnarray}
n_0 &=& 0.28665 \left(B_{60}\right)^{3/4}~{\rm fm^{-3}}~, 
\nonumber\\ 
\rho_0 &\equiv   & \epsilon_0/c^2 = 
4.2785\times 10^{17}~B_{60} {\rm\ kg~m^{-3}} \ .
\label{n0_rho0}
\end{eqnarray}
Strange matter would be the real ground state of matter at zero
pressure, only if
the energy per unit baryon number, at zero pressure, 
 $E_0\equiv \epsilon_0/n_0=
837.26 (B_{60})^{1/4}~{\rm MeV}$,  was below the energy per baryon
in the  maximally bound terrestrial nucleus, $^{56}{\rm Fe}$, equal
$930.4~{\rm MeV}$. This implies 
$B_{60}<1.525$. On the other hand, a consistent model should
not lead to a spontaneous fusion  of neutrons  into
droplets of u, d quarks (which would eventually transform into
droplets of strange matter - ``strangelets''). In view of the
fact, that $a_{\rm ud}/a=(1+2^{4/3})/3$, we get 
$E_0^{\rm (ud)}=1.127 E_0^{\rm (uds)}$. Therefore, the condition
of the stability of neutrons  with respect to a spontaneous
fusion into strangelets  $E_0^{\rm (ud)}=988.5(B_{60})^{1/4}~{\rm
MeV}>939.6~{\rm MeV}$, which implies $B_{60}>0.9821$. Summarizing, within
the simplest model of strange matter,   
 the bag constant
is constrained by $0.9821<B_{60}<1.525$.

An important quantity, relevant for the pulsations and stability of strange 
stars is the adiabatic index of strange matter, defined as 
$\gamma\equiv (n/P) {\rm d}P/{\rm d}n$. Dependence  of $\gamma$ on  baryon 
density of strange matter is shown  in Fig.~\ref{f:gamma}; it is 
qualitatively different from $\gamma(n)$ for ordinary neutron star matter. 
The values of 
$\gamma$ in the outer layers of strange stars are very large. Even at the 
highest densities, allowed for strange stars, the value of $\gamma$ is 
significantly higher, than the ultrarelativistic Fermi gas value $4/3$, 
predicted by the asymptotic freedom  of  QCD at $n\Longrightarrow \infty$. 

\section{Formulation of the problem and numerical code} \label{s:formul_code}
%
\subsection{Equations of stationary motion}

We refer to Bonazzola et al. (1993) for a complete description of the
set of equations to be solved to get general relativistic models of 
stationary rotating bodies. Let us simply recall here that, under the
hypothesis of stationarity, axisymmetry and purely azimuthal motion (no
convection), a coordinate system $(t,r,\theta,\varphi)$ can be chosen so
that the spacetime metric takes the form
\begin{eqnarray}
{\rm d}s^2 & = & -N^2 \, {\rm d}t^2 + B^2r^2\sin^2\theta 
({\rm d}\varphi - N^\varphi {\rm d}t)^2 
		\nonumber \\
  & &  + A^2 ({\rm d}r^2 + r^2 {\rm d}\theta^2) \label{e:metrique} \ ,
\end{eqnarray}
where $N$, $N^\varphi$, $A$ and $B$ are four functions of $(r,\theta)$. 
The Einstein equations result in a set of four elliptic equations for these
metric potentials\footnote{in this section $G=c=1$.}:
\begin{eqnarray}
& & \Delta_3 \, \nu \; = \; 
   4\pi A^2 ( E + 3 P + (E+P)U^2 ) \nonumber \\
	& & \qquad + \frac{B^2 r^2\sin^2\theta}{2N^2}(\partial N^{\varphi})^2 
         - \partial\nu \, \partial (\nu + \beta)   \label{e:Einstein1} \\
& & \tilde \Delta_3 \, (N^\varphi r\sin\theta)  \; = \; 
   - 16\pi \frac{NA^2}{B} (E+P) U  \nonumber \\
   & & \qquad 
	- r \sin\theta \, \partial N^\varphi  \, \partial (3\beta - \nu)
                                          \label{e:Einstein2}             \\
& & \Delta_2 \, [ (NB-1) \, r\sin\theta]  \; = \; 
    16 \pi NA^2 B P r \sin\theta
                                         \label{e:Einstein3}          \\
& & \Delta_2 \, (\nu+\alpha)    \; = \; 
  8\pi  A^2 \, [P + (E+P)U^2] \nonumber \\
    & &\qquad  + \frac{3B^2 r^2\sin^2\theta}{4N^2} (\partial N^{\varphi})^2 
	 - (\partial \nu)^2  \label{e:Einstein4}
\end{eqnarray}
where the following abreviations have been introduced:
\begin{equation}
\nu := \ln N\ , \quad \alpha :=\ln A \ , \quad \beta := \ln B \ , 
\end{equation}
\begin{eqnarray}
   \Delta_2  &:=&
    \ddps{}{r} + {1 \over r}\ddp{}{r} +{1 \over r^2}\ddps{}{\theta}
                               \\                             
   \Delta_3  &:=&
    \ddps{}{r} + {2 \over r} \ddp{}{r}
    +{1 \over r^2} \ddps{}{\theta} + {1 \over r^2 \tan\theta} \ddp{}{\theta}
                             \\                             
   \tilde \Delta_3  &:=&
     \Delta_3 
    - {1  \over r^2\sin^2\theta} \\
\partial a \, \partial b &:=& \ddp{a}{r} \ddp{b}{r}
    + {1 \over r^2} \ddp{a}{\theta} \ddp{b}{\theta} \ . 
\end{eqnarray}
In Eqs.~(\ref{e:Einstein1})-(\ref{e:Einstein4}), $U$ and $E$ 
are respectively the fluid 3-velocity and energy density, both
as measured by the locally non-rotating observer:
$U=Br\sin\theta(\Omega-N^\varphi)/N$, $E=\Gamma^2(\epsilon+P) - P$,
$\Gamma = (1-U^2)^{-1/2}$.

These equations are supplemented by the first integral of motion
\begin{equation}
  H + \nu  -\ln\Gamma = {\rm const.} \ ,		\label{e:int_prem}
\end{equation}
where $H$ is pseudo-enthalpy (or log-enthalpy) defined by 
\begin{equation}
H=\ln\left(
{\epsilon+P \over n E_0}
\right)~.
\label{e:H_def}
\end{equation}
%
The dimensionless EOS of strange matter can  be parametrized in terms of 
the $H$ variable as   
\begin{eqnarray}
\widetilde\epsilon &=& 3{\rm e}^{4H} + 1~,\nonumber\\
\widetilde P &=& {\rm e}^{4H} -1~.
\label{EOS_H}
\end{eqnarray}
Consequently, 
the baryon number density can expressed in terms of $H$ by 
\begin{equation}
n = n_0{\rm e}^{3H}~,
\label{n_H}
\end{equation}
where  for our model the baryon number density of strange
matter at zero pressure  $n_0=(3B/a)^{3/4}$. 

The equations of stationary motion involve the quantities
with dimensions of energy density  and powers of length, and
universal constants $G$ and $c$. These equations  can be
rewritten in a dimensionless form, if we introduce the
dimensionless quantities  $\widetilde\epsilon$, 
$\widetilde P$, $\widetilde r$, $\widetilde\Omega$, 
 via following relations, 
\begin{eqnarray}
\epsilon &=& B\widetilde\epsilon~, \nonumber\\
P &=&  B\widetilde P~,\nonumber\\
r &=& {c^2\over \sqrt{GB}}\widetilde r~,\nonumber\\
\Omega &=& {\sqrt{GB}\over c}\widetilde\Omega~.
\label{e:nondimvar}
\end{eqnarray}
 Notice, that the
solutions of dimensionless equations of stationary motion
will not depend explicitly on $B$. In order to recover  the solution in 
conventional units, for a specific value of $B$, one has just to 
use relations (\ref{e:nondimvar}). 

\subsection{Numerical procedure} \label{s:num_proc}

The non-linear elliptic equations (\ref{e:Einstein1})-(\ref{e:Einstein4})
are solved iteratively by means of the
multi-domain spectral method developed recently by Bonazzola et al. (1998b). 
In this method, the whole space is divided into three domains: D1: the
interior of the star, D2: an intermediate domain whose inner boundary is the
surface of the star and outer boundary a sphere located at 
$r:=R_{23}\sim 2 r_{\rm eq}$
(where $r_{\rm eq}$ is the equatorial coordinate-radius of the star),
and D3: the external domain whose inner boundary is the outer boundary of
D2 and which extends up to infinity, thanks to the compactification
$u=1/r$. A mapping $(\xi,\theta')\mapsto(r,\theta)$
satisfying $\theta=\theta'$ 
is introduced in each domain so that the domain boundaries lie at a constant
value of the coordinate $\xi$ (typically, $\xi=-1$, $0$ or $1$). 
Explicitly, this mapping reads
\begin{equation}
 \mbox{in D1:} \quad r = \alpha_1 \left[ \xi + (3\xi^4 - 2 \xi^6)
				\, F_1(\theta') \right] \ ,
			\quad \xi \in [0,1] \ , 		\label{e:map1}
\end{equation}
\begin{eqnarray}
  \mbox{in D2:} \quad r = \alpha_2 \left[ \xi - 1 
		+ {\xi^3-3\xi+2\over 4} F_2(\theta') \right] 
		+ R_{23} \ ,  & & \nonumber \\
	  \xi \in [-1,1] \ , & &  				\label{e:map2}
\end{eqnarray}
\begin{equation}
   \mbox{in D3:} \quad u := {1\over r} = {1\over 2 R_{23}} (1-\xi) \ , 
			\quad \xi \in [-1,1] \ ,		\label{e:map3}
\end{equation}
where the functions $F_1(\theta')$ and $F_2(\theta')$ are related to 
the equation $r=S(\theta)$ of the stellar surface by
$S(\theta) = \alpha_1[1+F_1(\theta)] = \alpha_2[-2+F_2(\theta)]+R_{23}$.
The mapping (\ref{e:map1})-(\ref{e:map3}) is a specialization to the
axisymmetric case of the 3-D mapping introduced in Bonazzola et al. (1998b),
to which the reader is referred for more details. 

A spectral expansion of each relevant field is then performed with respect
to the coordinates $(\xi,\theta')$. We use Chebyshev polynomials
in $\xi$ and trigonometric polynomials or Legendre polynomials in $\theta'$
(see Bonazzola et al. 1999 for the use of spectral methods in relativistic
astrophysics). 

Since the discontinuities in the physical fields (strong discontinuity in
the density, cf. Fig.~\ref{f:prof_dens1} below, 
discontinuity in the second derivative for the metric potentials)
are located at the boundary between two domains (namely D1 and D2), 
the applied spectral method is free from any Gibbs phenomenon and leads to
a very high precision. This is illustrated in Fig.~5 of Bonazzola et al. (1998b)
which concerns the case of a rotating constant-density body 
(hence with a strong  discontinuity of the density field at the surface) in 
the Newtonian theory. An analytical solution is available in this case
(MacLaurin ellipsoid) and can be used to evaluate the accuracy of the code.
The code gives a rapidly rotating MacLaurin ellipsoid with a relative error
of the order $10^{-12}$ with 49 (resp. 25) degrees of freedom in $r$ 
(resp. $\theta$).  Figure~5 of Bonazzola et al. (1998b) also demonstrates that
the error is {\em evanescent}, i.e. that it decreases exponentially with
the number of degrees of freedom (i.e. grid points). 

This represents a major improvement with respect to the spectral method
developed previously by Bonazzola et al. (1993) to compute 
stationary configurations
of rotating bodies in general relativity. 
This method has been used to compute
rapidly rotating neutron star models by Salgado et al. (1994a,b),
Haensel et al.~(1995), Goussard et al. (1997, 1998) 
and Nozawa et al. (1998) but it 
could not have been employed as such to compute models of rotating strange
stars: the Gibbs phenomenon induced by the density discontinuity at the
stellar surface would have been too large. 

\begin{figure}
\centerline{\includegraphics[width=8.5cm]{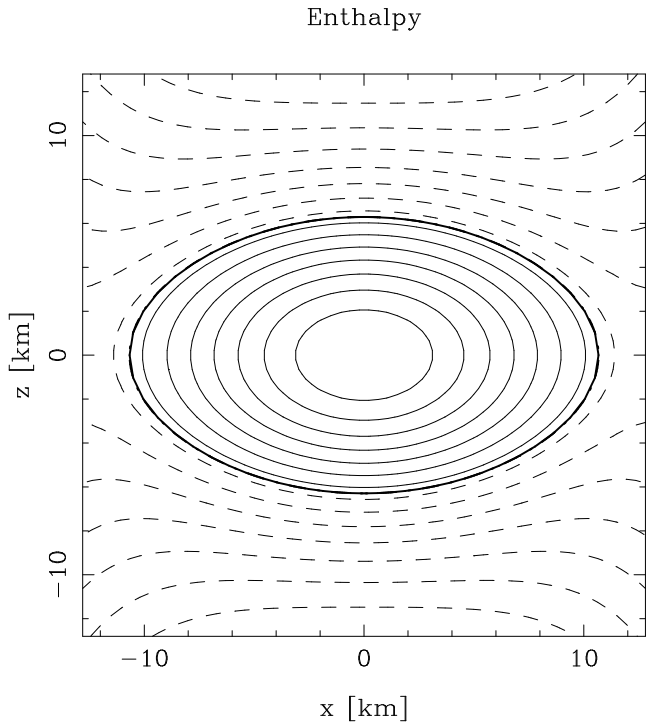}}
\caption[]{\label{f:coupe1}
Meridional plane cross section of a rapidly rotating strange star, 
for $B=60~{\rm MeV~fm^{-3}}$. The
star's baryon mass is $M_{\rm B} = 2.00\, {\rm M}_\odot$, gravitational mass
$M=1.60\, M_\odot$, and rotation period $P=0.87{\ \rm ms}$. 
The coordinates $x$ and $z$ are defined from the coordinates 
$(r,\theta,\varphi)$ introduced in the line element (\ref{e:metrique})
by $x=r\sin\theta\cos\varphi$ and $z=r\cos\theta$. 
The rotation axis is the $z$-axis. 
The various lines are isocontours of the log-enthalpy $H$. 
Solid lines indicate a positive value of $H$ and dashed lines a negative
value. In this latter case (vacuum), 
$H$ is defined from Eq.~(\ref{e:int_prem}). The thick solid line denotes
the stellar surface.}
\end{figure}

\begin{figure}
\centerline{\includegraphics[width=8.5cm]{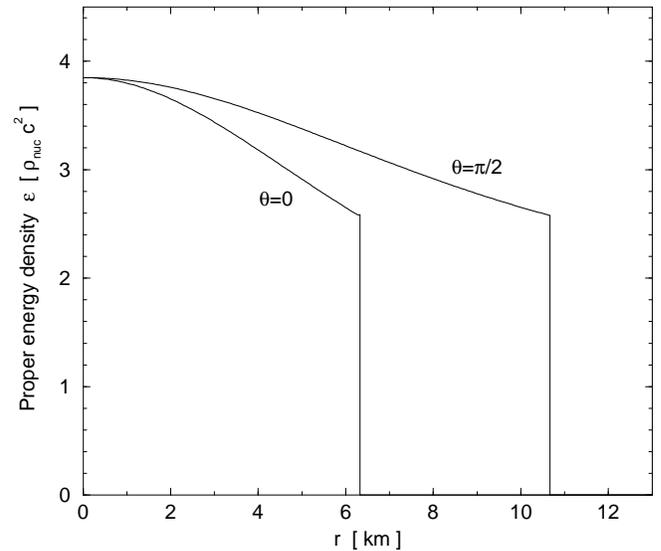}}
\caption[]{\label{f:prof_dens1}
Profiles of the proper energy density $\epsilon$ in the equatorial plane
($\theta=\pi/2$) and along the rotation axis ($\theta=0$) for the
rotating strange star model shown in Fig.~\ref{f:coupe1}.}
\end{figure}

\subsection{Description of a typical model}

A rapidly rotating strange star model obtained from the code is 
shown in Figs.~\ref{f:coupe1} and \ref{f:prof_dens1}. Assuming
$B=60{\rm\ MeV\, fm}^{-3}$, this model has
a baryon mass $M_{\rm B} = 2.00\, {\rm M}_\odot$, a gravitational mass
$M=1.60\, {\rm M}_\odot$, and a rotation period $P=0.87{\ \rm ms}$.  
 The central
baryon density is $n_{\rm c} = 0.42 {\rm\ fm}^{-3}$, the central values
of the metric potentials $N$, $A$, $B$ and $N^\varphi$ 
[cf. Eq.~(\ref{e:metrique})] are respectively 0.65, 1.47, 1.47 and
$0.46\, \Omega$. 
The boundary between the computational domains D1 and D2 introduced in
Sect.~\ref{s:num_proc} coincides with the stellar surface (thick solid line
in Fig.~\ref{f:coupe1}). The boundary between the computational domains D2 
and D3 is out of the scope of the Figure. The energy density profile
corresponding to this model is shown in Fig.~\ref{f:prof_dens1}. Note the
strong discontinuity at the stellar surface: the jump in density
is $\sim 2/3$ of the central density. 

\begin{figure}
\centerline{\includegraphics[width=8.5cm]{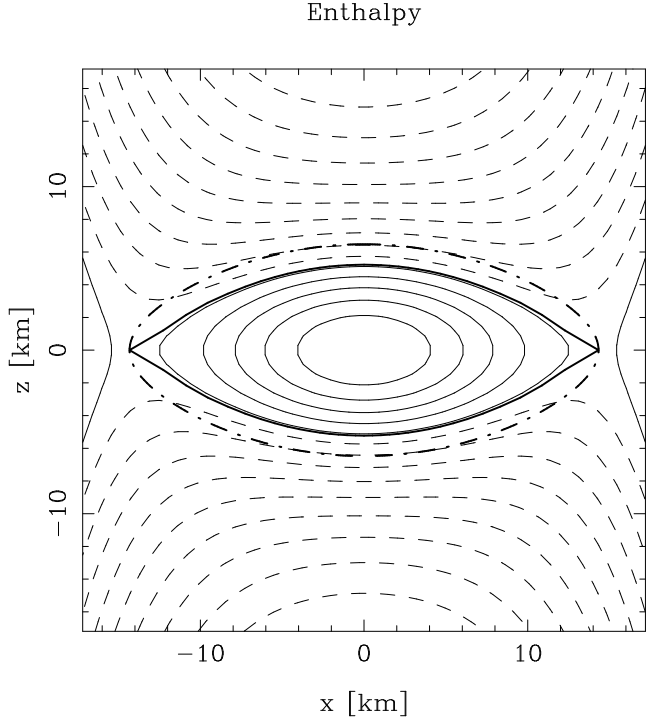}}
\caption[]{\label{f:coupe_kep}
Same as Fig.~\ref{f:coupe1} but for a strange star rotating at the Keplerian
limit. The
star's baryon mass is $M_{\rm B} = 2.34\, {\rm M}_\odot$, gravitational mass
$M=1.89\, {\rm M}_\odot$, and rotation period $P=0.82{\ \rm ms}$. 
Thin dashed and solid lines are isocontours of the log-enthalpy $H$.   
Thin dashed lines correspond to  negative values, and  thin solid 
lines to  positive values of $H$, respectively.  
The thick dot-dashed line denotes the boundary between the computational
domains D1 and D2.}  
\end{figure}

\subsection{Treatment of configurations at the Keplerian limit}
\label{s:trait_kep}

The maximum rotation rate that a star can sustain is reached when the
velocity at the equator equals the velocity of an orbiting particle
(a higher velocity would result in a centrifugal break-up of the star). 
This limit is called the {\em Keplerian limit}. It is reached when the
derivative $\partial H/\partial r$ vanishes at the stellar equator. 
The surface of the star is then no longer smooth but exhibits a cusp
along the equator (see Fig.~\ref{f:coupe_kep}). Such a non-differentiable
surface cannot be decribed by the mapping (\ref{e:map1})-(\ref{e:map3})
because the functions $F_1(\theta)$ and $F_2(\theta)$ are assumed to 
be expandable in $\cos(k\theta)$ series (cf. Bonazzola et al. (1998b)), 
which implies that they are smooth functions of $\theta$. 
The solution to this problem consists in freezing the adaptation 
of the mapping to the stellar surface  when
the ratio 
$(\partial H/\partial r)_{\rm eq}\; / \; (\partial H/\partial r)_{\rm pole}$
passes below a certain threshold
during the iteration process. For instance, this threshold was chosen 
to be $0.28$ in the computation shown in Fig.~\ref{f:coupe_kep}. 
Consequently, in the final result, the density discontinuity at the
stellar surface no longer coincides with the boundary between 
the computational domains D1 and D2, except at the equator. 
In this case, a Gibbs phenomenon is present. The accuracy of the calculation
is then lower than when the mapping is adapted to the surface of the star. 
However, since the stellar interior covers most of domain D1
(cf. Fig.~\ref{f:coupe_kep}), the Gibbs phenomenon is rather limited.
It is notably less severe than if D1 would have been
a sphere of radius $r_{\rm eq}$, as in our previous numerical method
(Bonazzola et al.~1993). 

\subsection{Tests of the numerical code} \label{s:tests}

Various tests have been passed by the code. 
First of all, we recover previous results, e.g. those presented
in Nozawa et al. (1998), when using an EOS for neutron star
matter instead of the strange quark matter EOS presented in 
Sect.~\ref{s:EOS}. 

Regarding the treatment of the density discontinuity at the stellar surface,
the code has been tested on the computation of MacLaurin ellipsoids
(homogeneous rotating bodies in the Newtonian regime), and gives 
excellent results, as recalled in Sect.~\ref{s:num_proc}:
the relative accuracy reaches $10^{-12}$ !

Another type of test is the evaluation of the 
virial identities GRV2 (Bonazzola 1973, Bonazzola \& Gourgoulhon 1994)
and GRV3 (Gourgoulhon \& Bonazzola 1994), this latter being  a 
relativistic generalization of the classical virial theorem.
GRV2 and GRV3 are integral identities
which must be satisfied by any solution of the Einstein 
equations (\ref{e:Einstein1})-(\ref{e:Einstein4})
and which are not imposed during the numerical procedure
(cf. Nozawa et al. 1998 for details on the computation of GRV2 and GRV3). 
When presenting numerical results, we will systematically
give the accuracy by which the numerical solution satisfies these virial
identities. 

As discussed in Sect.~\ref{s:trait_kep}, the computation of configurations
rotating at the Keplerian limit is not free of Gibbs phenomenon. In order
to gauge the resulting numerical error, we performed various computations
of the same Keplerian configuration, by varying the numbers of 
coefficients in the spectral method and by varying the threshold on
$(\partial H/\partial r)_{\rm eq}\; / \; (\partial H/\partial r)_{\rm pole}$
for freezing the mapping. We systematically used three sets of numbers
of coefficients (= number of collocation points): $(N_r,N_\theta)$ = 
$(33,17)$, $(37,19)$ and $(41,21)$ in each of the three domains. 
The threshold on 
$(\partial H/\partial r)_{\rm eq}\; / \; (\partial H/\partial r)_{\rm pole}$
was varied from $0.28$ down to $0.20$. 
The variation of all these parameters lead to a relative change of
the numerical solution of the order of or below $1\%$. 
This gives an estimation 
of the error of our method in computing strange stars at the Keplerian
limit. Note the GRV2 and GRV3 errors for Keplerian configurations revealed
to be better than this. For instance, for the configuration shown
in Fig.~\ref{f:coupe_kep}, the GRV2 (resp. GRV3) error is 
$3\, 10^{-5}$ (resp. $1.3\, 10^{-3}$).

\begin{figure}
\centerline{\includegraphics[width=8.5cm]{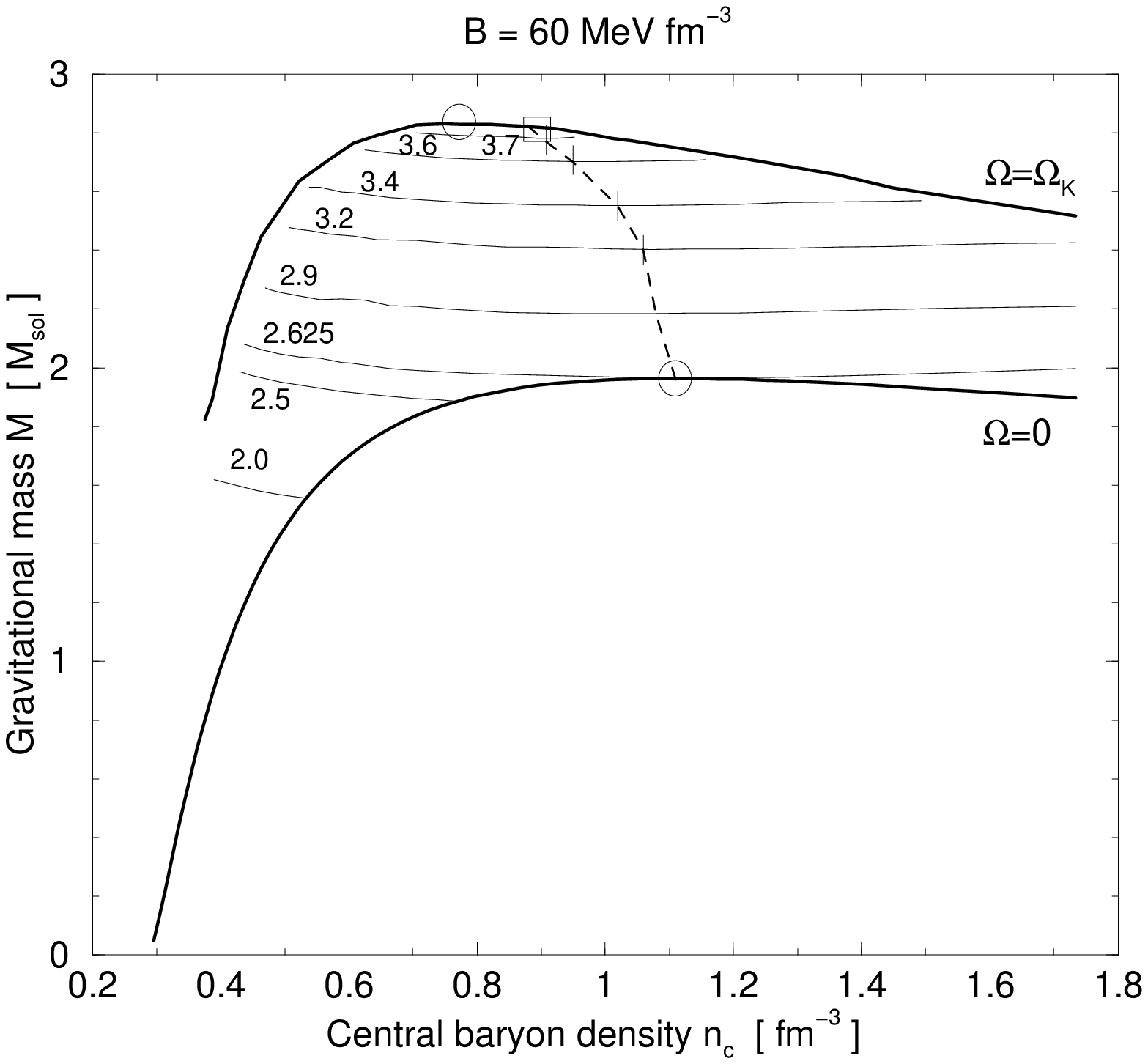}}
\caption[]{\label{f:mass_densc}
Rotating strange star configurations in the mass - central density plane. 
The static limit and Keplerian limit are denoted by thick solid lines. 
Thin solid lines denotes sequences of constant baryon number; they are
labelled by the baryon mass (in ${\rm M}_\odot$ units). The dashed line marks
the limit of stability with respect to axisymmetric pertubations; 
configuration to the right of the dashed line are (secularly) unstable.
Its intersection with the Keplerian limit curve is marked by an open square.
It is the location of the fastest rotating stable star. 
Maximum mass configurations are denoted by circles}
\end{figure}

\section{Numerical results} \label{s:result}

\subsection{General properties} \label{s:present}

For rigid rotation, the equilibrium configurations span a two-dimensional
domain. This domain is shown in Fig.~\ref{f:mass_densc} in the 
central density - gravitational mass plane.
Selected configurations are listed in Table~\ref{t:models}. 
The symbols have the following meaning: $M$ is the gravitational mass;
$M_{\rm B}$ is the baryon mass; $\Omega$ is the angular velocity,
$P$ the corresponding rotation period; $n_{\rm c}$ is the central baryon density;
$\rho_{\rm c}$ is the central proper energy density 
divided by $c^2$; $H_{\rm c}$ is the central
log-enthalpy [Eq.~(\ref{e:H_def})]; $R_{\rm eq}$ is the circumferential radius,
i.e. the length of the equator 
[as given by the metric (\ref{e:metrique})] divided
by $2\pi$; $r_{\rm eq}$ is the equator $r$ coordinate;  $r_{\rm pole}/r_{\rm eq}$
is the coordinate oblateness of the star; $I$ is the moment of inertia (defined
as $J/\Omega$); $J$ is the star angular momentum; $Q$ is the
quadrupole moment defined as in Salgado et al. (1994a) or 
Laarakkers \& Poisson (1999);
$T/W$ is the ``kinetic to gravitational
energy'' ratio (see Sect.~\ref{s:bar_instab}); 
$U_{\rm eq}$ is the rotation velocity at the equator
as measured by a locally non-rotating observer (those observer whose 4-velocity
is the normal to the $t={\rm const}$ hypersurfaces); $z_{\rm eq}^{\rm f}$
is the redshift for an emission at the equator and in the direction of
rotation, $z_{\rm eq}^{\rm b}$ the redshift for an emission at the equator 
and in the direction opposite to rotation, $z_{\rm pole}$ the redshift at the
stellar pole; $N_{\rm c}$, $N^{\varphi}_{\rm c}$, $A_{\rm c}$ and $B_{\rm c}$
are the metric potentials [Eq.~(\ref{e:metrique})] at the stellar center.
Finally GRV2 and GRV3 are the two virial error indicators discussed in 
Sect.~\ref{s:tests}.

\begin{table*}
\caption{\label{t:models} Selected strange star models 
($B=60 {\rm\  MeV\, fm}^{-3}$); symbols are defined in Sect.~\ref{s:present}.}
\begin{tabular}{llllll}
\hline
Model & $M=1.4\, {\rm M}_\odot$ & 
$M_{\rm max}^{\rm stat}$ & $M=1.4\, {\rm M}_\odot$ 
					& $M_{\rm max}^{\rm rot}$ & $\Omega_{\rm max}$ \\
      &  $\Omega=0$   &  & $P=1.56{\rm\ ms}$ & & \\
\hline
$M$ [${\rm M}_\odot$] & 1.400 & 1.964 & 1.400 & 2.831 & 2.822  \\ 
$M_{\rm B}$ [${\rm M}_\odot$] & 1.778 & 2.625 & 1.767 & 3.751 & 3.756\\
$\Omega$ [${\rm rad\, s}^{-1}$] & 0 & 0 & 4028. & 9547. & 9916. \\
$P$ [ms] & - & - & 1.560 & 0.6581 & 0.6337 \\
$n_{\rm c}$ [${\rm fm}^{-3}$] & 0.4842 & 1.110 & 0.4615 & 0.7486 & 0.8698 \\
$\rho_{\rm c}$ [$10^{18} {\rm\ kg\, m}^{-3}$] & 0.7524 & 2.059 & 0.7125 & 1.261 & 1.517 \\
$H_{\rm c}$ &  0.1747 & 0.4514 & 0.1588 & 0.320 & 0.370 \\
$R_{\rm eq}$ [km] & 10.77 & 10.71 & 11.19 & 16.54 & 16.02 \\
$r_{\rm eq}$ [km] & 8.573 & 7.532 & 8.946 & 11.37 & 10.86 \\
$r_{\rm pole}/r_{\rm eq}$ & 1. &  1. & 0.8935 & 0.4618 & 0.4762 \\
$I$ [$10^{38} {\rm\ kg\, m}^2$] & 1.492 & 2.155 & 1.603 & 6.534 & 6.101 \\
$J$ $[G{\rm M}_\odot^2/c]$ & 0 & 0 & 0.7332 & 7.084 & 6.871 \\
$cJ/(GM^2)$ & 0 & 0 & 0.3741 & 0.8837 & 0.8625 \\
$Q/(M R_{\rm eq}^2)$ & 0 & 0 & 0.0250 & 0.0749 &  0.0713 \\
$c^4 Q / (G^2 M^3)$ & 0 & 0 & 0.7305 & 1.172 & 1.054 \\
$T/W$ & 0 & 0 & 0.0348 & 0.2100 & 0.2007 \\
$U_{\rm eq}$ [c] & 0 & 0 & 0.1565 & 0.5731 & 0.5812 \\
$z_{\rm eq}^{\rm f}$ & 0.2742 & 0.4768 & 0.0463 & $-0.3608$ & $-0.3634$ \\
$z_{\rm eq}^{\rm b}$ & 0.2742 & 0.4768 & 0.5351 &  2.584 & 2.722 \\
$z_{\rm pole}$ & 0.2742 & 0.4768 & 0.2825 & 0.8070 & 0.847 \\
$N_{\rm c}$ & 0.6590 & 0.4312 & 0.6653 & 0.4019 & 0.374 \\
$N^{\varphi}_{\rm c}/\Omega$ & 0 & 0 & 0.4203 & 0.7320 & 0.7540 \\
$A_{\rm c}=B_{\rm c}$ & 1.439 & 1.914 & 1.432 & 2.135 & 2.234 \\
\hline 
GRV2 & $1.\; 10^{-12}$ & $9.\; 10^{-12}$ & $1.\; 10^{-4}$ & $2.\; 10^{-4}$ & $9.\; 10^{-4}$ \\
GRV3 & $2.\; 10^{-12}$ & $1.\; 10^{-11}$ & $2.\; 10^{-4}$ & $1.\; 10^{-4}$ & $2.\; 10^{-3}$ \\
\hline                                                           
\end{tabular}
\end{table*}

\begin{figure}
\centerline{\includegraphics[width=8.5cm]{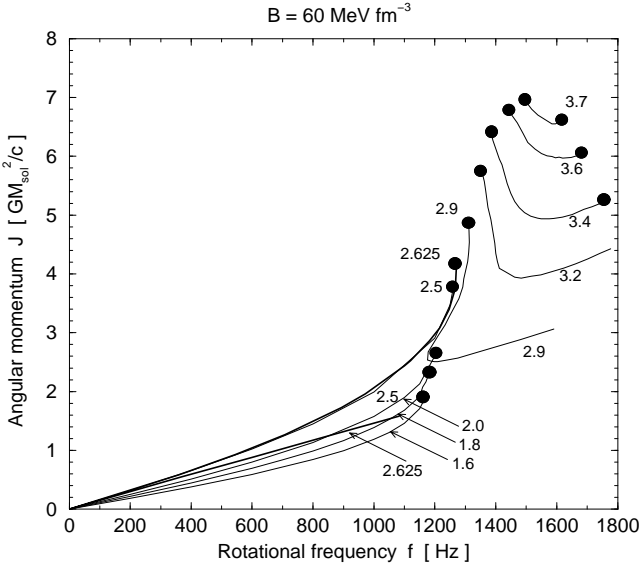}}
\caption[]{\label{f:mom_cin}
Angular momentum as a function of the rotation frequency 
$f=\Omega/(2\pi)$ along sequences
of constant baryon number. Each sequence is labelled by its baryon mass in
${\rm M}_\odot$ units. The heavy black circles denotes Keplerian configurations.}
\end{figure}

\subsection{Evolutionary sequences} \label{s:evol}

A strange star which slowly loses energy and angular momentum via 
electromagnetic
or gravitational radiation keeps its
 total baryon number constant. Therefore, we
can compute evolutionary sequences of strange stars as sequences at fixed
baryon mass $M_{\rm B}$. Similarly to neutron stars, two categories
of strange stars can be distinguished: {\em normal} stars, which have a
baryon mass lower than the maximum baryon mass of static configurations
$M_{\rm B,max}^{\rm stat} = 2.625 \, B_{60}^{-1/2} \; {\rm M}_\odot$ 
(cf. Table~\ref{t:models}), and {\em supramassive} stars, which have a
baryon mass greater than $M_{\rm B,max}^{\rm stat}$. Any normal star belongs
to an evolutionary sequence which terminates at a static configuration. On
the contrary, supramassive stars exist only by virtue of rotation. The
two families clearly appear in Fig.~\ref{f:mom_cin}, which is a plot of the
angular momentum as a function of the rotation frequency for evolutionary
sequences. The supramassive sequences are not connected with the static
limit $\Omega=0$. At both ends, they terminate by Keplerian configuration. 
The limiting case $M_{\rm B} = M_{\rm B,max}^{\rm stat}$ is denoted by
the thick line in Fig.~\ref{f:mom_cin}. 

A characteristic feature of some supramassive sequences also appears 
in Fig.~\ref{f:mom_cin}: they can be spun-up by angular momentum loss:
$d\Omega/dJ <0$. This effect is well known for Newtonian stars with soft EOS
(adiabatic index close to $4/3$) (Shapiro et al. 1990), as well as for
(relativistic) neutron stars (Cook et al. 1994, Salgado et al 1994a).
As can be seen from Fig.~\ref{f:mom_cin}, this effect is very prononced 
for strange stars. 

Beside the fact that they can represent evolution path of rotating strange
stars, another motivation for computing constant baryon number configurations
is that this permits a stability analysis, as discussed in 
Sect.~\ref{s:max_rot_stable} below.

\subsection{Comparison with previous works} \label{s:compar}

First of all, in the
non-rotating case, we recover the results presented in Haensel et al.~(1986)
for massless strange quarks [compare the $M_{\rm max}^{\rm stat}$ model in 
Table~\ref{t:models} with Eq.~(28) of Haensel et al.~(1986)].

The first rapidly rotating model of a strange star has been computed by
Friedman (unpublished, quoted in Glendenning (1989a,b)). The obtained
maximum angular velocity is $6\%$ lower than ours 
(last column in Table~\ref{t:models})\footnote{Friedman's result is
presented for ${\tilde B}^{1/4} := (\hbar c)^{3/4} B^{1/4} = 170{\rm\ MeV}$,
which corresponds to $B=108.7{\rm\ MeV\, fm}^{-3}$. It must be rescaled
to $B=60{\rm\ MeV\, fm^{-3}}$, according to the law (\ref{e:nondimvar}),
in order to be compared with results listed in Table~\ref{t:models}.}.
The corresponding gravitational mass is $7\%$ lower than ours, and the
central energy density is $7\%$ greater than ours. 

The only other rapidly rotating model of strange star published in the 
literature is that of Lattimer et al.~(1990). The maximum angular velocity 
obtained by these authors is 
$2\%$ lower than ours (last column in Table~\ref{t:models})
\footnote{Lattimer et al.~(1990) results (their Table~7) is presented
for $B=1.634\times 56 = 91.5 {\rm\ MeV\, fm^{-3}}$. It must be rescaled
to $B=60{\rm\ MeV\, fm^{-3}}$, according to the law (\ref{e:nondimvar}),
in order to be compared with results listed in Table~\ref{t:models}.}.
The corresponding mass is $12\%$ lower than ours, 
the circumferential radius $R_{\rm eq}$ $14\%$ lower,
the ratio $T/W$ $10\%$ lower, and the polar redshift (linked to the
injection energy $\beta$ given by Lattimer et al. (1990)  by 
$z_{\rm pole} = \beta^{-1/2}-1$) $15\%$ lower. The largest discrepancy 
arises for the baryon mass $M_{\rm B}$: 
the value reported by Lattimer et al. (1990) is
$21\%$ lower than ours. However we suspect some error in the presentation
of the results by Lattimer et al. (1990), because (i) their ratio
$M_{\rm B}/M$ is only $1.18$, whereas it is as large as $1.34$ for the 
non-rotating maximum mass model (for comparison, our value of $M_{\rm B}/M$
for the $\Omega_{\rm max}$ model is $1.35$), (ii) in the neutron star case,
large discrepancies with baryon masses presented in Lattimer et al. (1990) 
have been noticed previously by Salgado et al. (1994a)
[see Sect.~5.1.2 of Salgado et al. (1994a)]. Note that the numerical 
code employed by Salgado et al. (1994a) has been successfully compared
with codes from two other groups in Nozawa et al. (1998) and did not 
show any trouble with the baryon mass. Unfortunately, the baryon mass
of Friedman's model is not reported in Glendenning (1989a,b). 

Apart from the baryon mass problem discussed above, the discrepancy with
Friedman and Lattimer et al. (1990) results remains rather large. It may be
explained as follows. Both Friedman's code [presented in Sect.~II.a of 
Friedman et al.~(1986)] and Lattimer et al. (1990) code employ the
Butterworth \& Ipser (1976) finite difference technique for 
numerically solving the partial differential equations resulting 
from Einstein's equations. Butterworth \& Ipser (1976) considered explicitly
the case of a strong discontinuity in the density field at the stellar
surface [their Sect.~III.c)ii)]. However they treat it only along the
radial direction (by using a modified Lagrangian-polynomial fit in $r$) and
keep a Legendre expansion for the angular variable $\mu:=\cos\theta$. This
is correct as long as
the stellar surface is exactly spherical, i.e. the star is
static. But when the surface deviates from spherical symmetry (as is the case 
for rapidly rotating stars, cf. Figs.~\ref{f:coupe1} and \ref{f:coupe_kep}),
there is also a discontinuity in the $\theta$ direction. This discontinuity,
when described by means of Legendre polynomials, which are smooth functions,
inevitably generates spurious oscillations --- the so-called Gibbs phenomenon. 
We believe that it is this Gibbs phenomenon which explains the 
discrepancy between Friedman and Lattimer et al. (1990) results and ours.

\section{Maximally rotating configuration and scaling with $B$}  
\label{s:rotmax}

\subsection{Maximally rotating stable configuration} \label{s:max_rot_stable}

A natural bound on $\Omega$  results from the mass shedding  condition,
corresponding to the Keplerian angular velocity, $\Omega_{\rm K}$ 
discussed in Sect.~\ref{s:trait_kep}:
stationary configurations of uniformly rotating strange
stars with $\Omega>\Omega_{\rm K}$ do not exist. We will also
require, that the stationary, uniformly rotating configuration
be stable with respect to axisymmetric perturbations
(see e.g. Sect.~2.5.1 of Friedman 1998). We assume that 
the strange star is sufficiently hot, so that it is not subject to 
the r-mode instabilities (Madsen 1998). 
The maximum value of $\Omega$ for stationary, uniformly rotating models of 
strange stars which satisfy these two conditions, will be denoted by 
$\Omega_{\rm max}$, and the corresponding configuration will be 
called a {\it maximally rotating} one. 

The stability with respect to axisymmetric perturbations can be investigated
by means of the turning point theorem established by Friedman et al. (1988).
We use the following variant of it [also used by Baumgarte et al. (1998)]: 
along an evolutionary
sequence (i.e. a sequence at constant baryon number, Sect.~\ref{s:evol}),
the change in stability is reached when the gravitational mass reaches
a minimum. Using this criterion, we have determined the boundary 
separating  
stable and unstable configurations by linking the points of mimimum $M$ along
the constant baryon mass sequences in Fig.~\ref{f:mass_densc}: the 
configurations located to the left (resp. to  the right) 
of the dashed line in Fig.~\ref{f:mass_densc}
are stable (resp. unstable) with respect to axisymmetric perturbations. 
Moreover, the intersection of this line with the $\Omega=\Omega_{\rm K}$ line
in Fig.~\ref{f:mass_densc} marks the maximum angular velocity allowable
for stable rotating strange stars, i.e. the $\Omega_{\rm max}$ configuration. 
It is denoted by a square and is listed in the last column of 
Table~\ref{t:models}. 

As noted by Cook et al. (1994) and Stergioulas \& Friedman (1995), the
$\Omega_{\rm max}$ configuration and the configuration with 
maximum gravitational mass, $M_{\rm max}^{\rm rot}$, do not coincide, 
although they are close to each other for typical neutron star models. 
We also found this for rotating strange stars, and the effect is more
pronounced than in the case of neutron stars: it can be seen easily
in Fig.~\ref{f:mass_densc}. The parameters of the 
$M_{\rm max}^{\rm rot}$ and
$\Omega_{\rm max}$ configurations are listed in  the last two columns in
Table~\ref{t:models}. $\Omega_{\rm max}$ is $4\%$ higher than 
$\Omega(M_{\rm max}^{\rm rot})$. 
Note that the $M_{\rm max}^{\rm rot}$ configuration is on the stable side,
so that the situation is similar to that presented in Fig.~9 of 
Stergioulas \& Friedman (1995), or Fig.~8 of Koranda et al. (1997).

\subsection{Formul\ae\ for $\Omega_{\rm max}$}

The scaling properties of the equation of the 
stationary motion
imply that  $\Omega_{\rm max}\propto B^{1/2}$. The calculation
of $\Omega_{\rm max}$ involves locating the threshold for the
instability with respect to the axisymmetric perturbations, and
requires therefore a higly precise numerical code for the 2-D
calculations of stationary configurations. Therefore, the
precision  with which the scaling formula for $\Omega_{\rm max}$
is fulfilled reflects the overall precision of our numerical
calculations. Our numerical results for $\Omega_{\rm max}$ 
 can  summarized in  {\it exact} formulae
\begin{equation}
\Omega_{\rm max}=9.92\; 10^3 ~\sqrt{B_{60}} {\rm\  rad\, s^{-1}}~,~~
P_{\rm min}=
{
0.634\over
 \sqrt{B_{60}}
} ~{\rm ms}~. 
\label{Omega_B}
\end{equation}
We checked that 
the scaling $\Omega_{\rm max}
\propto \sqrt{B}$ holds  with a very high precision. 
 
Within our simple model of EOS of strange matter, the constraints on the 
bag constant $B$, Sect. 2, imply 
\begin{equation}
 0.513~{\rm ms} < P_{\rm min} <0.640~{\rm ms}~. 
\label{Pmin.limits}
\end{equation}

In the case of dense baryon matter, exact results for
 the maximum rotation frequency of uniformly rotating neutron
star models, for a very broad set of realistic {\it causal} EOS,
show an interesting correlation with values of the mass and
radius of the {\it static} configuration with maximum allowable
mass, $M_{\rm max}^{{\rm stat}}$, $R_{M_{\rm max}}^{{\rm
stat}}$ (Haensel \& Zdunik 1989, Shapiro et al. 1989, 
Friedman et al. 1989, Friedman 1989, Haensel et al. 1995). 
The most recent form of
such an  ``empirical formula'' for $\Omega_{\rm max}$, derived
in (Haensel et al. 1995), reads
\begin{eqnarray}
\Omega_{\rm max}&\simeq&C_{\rm NS}
\left( 
{M_{\rm max}^{\rm stat}\over {\rm M}_\odot}
\right)^{1\over 2}
\left(
 {R_{M_{\rm max}}^{\rm stat}\over 10~{\rm km}}
 \right)^{-{3\over 2}}~,\nonumber\\
C_{\rm NS}&=&7730 {\rm\  rad\, s^{-1}}~,
\label{empNS}
\end{eqnarray}
where the prefactor 
$C_{\rm NS}$, which {\it does not depend} on the EOS of dense
baryon matter,   has been obtained via fitting  
``empirical formula'' to exact numerical results for realistic,
causal EOS of baryon matter. The
``empirical formula'' reproduces exact results with a
surprisingly high precision, the relative deviations from Eq.
(\ref{empNS}) not exceeding 5\%. 
In the case of strange stars, built of strange matter of
massless quarks, the formula of the type (\ref{empNS}) is
{\it exact}, albeit with a different numerical prefactor. 
Using our numerical results, we get 
 $C_{\rm SS}= 7.84~10^3 ~{{\rm rad}\over {\rm s}}$, 
so that $C_{\rm SS}\simeq  C_{\rm NS}$ within better 
than 2\%. Therefore, ``empirical formula'', derived originally for 
 (ordinary) neutron stars, holds also, with very good 
precision, for strange stars. Our conclusion agrees with unpublished result 
of Friedman (quoted in Glendenning 1989a), and with Lattimer et al. (1990).  

\subsection{Formul\ae\ for maximum mass configurations}

Rotation increases maximum allowable mass of strange stars, and 
the equatorial radius of the maximum mass configuration. 
Our results for the maximum mass of rotating configurations, 
$M_{\rm max}^{\rm rot}$, 
and its equatorial radius,
$R_{{\rm eq},M_{\rm max}^{\rm rot}}$,
can be summarized in two exact formulae,
\begin{eqnarray}
M_{\rm max}^{\rm rot}&=&{2.831\over \sqrt{B_{60}}}~{\rm M}_\odot~,
\nonumber\\
R_{{\rm eq},M_{\rm max}}^{\rm rot}&=&{16.54\over \sqrt{B_{60}}}~{\rm km} \,
\label{MR.rot}
\end{eqnarray}
The ratio
of the maximum mass of rotating configurations  to that of  static 
ones, and the ratio of the  corresponding equatorial radii,  are thus  
independent of $B$, 
\begin{equation}
{M^{\rm rot}_{\rm max}\over M^{\rm stat}_{\rm max}} = 1.44~,~~
{R_{{\rm eq},M_{\rm max}}^{\rm rot}\over 
R_{M_{\rm max}}^{\rm stat}} = 1.54~.
\label{RMmax.rel}
\end{equation}

It has been pointed out by Lasota et al. (1996), that for realistic causal 
baryonic EOS the ratios 
$M^{\rm rot}_{\rm max}/
 M^{\rm stat}_{\rm max}$ and 
$R_{{\rm eq},M_{\rm max}}^{\rm rot}/
 R_{M_{\rm max}}^{\rm stat}$ 
are very weakly dependent on the EOS, and can be very well approximated 
(within a few percent) by constants $C_M=1.18$ and $C_R=1.34$, 
respectively. 

In the case of strange stars the values of $C_M$ and $C_R$ are significantly 
higher; rotation increases the value of $M_{\rm max}$ by 44\% compared 
to 18\% for neutron stars, while the corresponding increase in the equatorial 
radius is 54\% compared to 34\% for neutron stars. 

We conclude that``empirical formula'' for $\Omega_{\rm max}$ has more universal
character than the ``empirical formulae'' for $M_{\rm max}^{\rm rot}$, 
$R_{{\rm eq},M_{\rm max}}^{\rm rot}$,
 proposed in Lasota et al. (1996); the former 
applies both for neutron  and strange stars, while the latter 
describe neutron  stars only.

\section{About triaxial instabilities of rapidly rotating strange stars}
\label{s:bar_instab}

\subsection{Known results on the threshold in $T/W$ for triaxial 
instabilities}

Rapidly rotating neutron stars can be subject to a spontaneous breaking 
of their symmetry about the rotation axis, resulting from
some triaxial instability, when their rotation rate exceeds a certain
threshold (see e.g. Bonazzola \& Gourgoulhon (1997),
Friedman (1998) or Stergioulas (1998) for a review). 
In order for the symmetry breaking to take place, a (secular) dissipative
mechanism has to be working. Basically two such mechanisms can be contemplated:
(i) viscosity and (ii) coupling with gravitational radiation 
(Chandrasekhar-Friedman-Schutz (CFS) instability).
For compact stars that are accelerated by accretion in X-ray binary systems,
these triaxial instabilities may be an important source of gravitational
waves, in the frequency band of the LIGO/VIRGO interferometric detectors  
currently under construction. Another situation where these triaxial
instabilities might develop is in compact stars newly formed after a stellar
core gravitational collapse. 

In the present paper, we have restricted ourselves to stationary and 
axisymmetric rotating strange star models. Therefore, we 
could not  compute the triaxial instability threshold of rotating strange
stars. 
 However, an indicator of the stability of a
rotating self-gravitating body is its kinetic to gravitational potential 
energy ratio $T/W$, which can be computed for any equilibrium configuration. 
For instance, it is well known that a homogeneous Newtonian rotating 
body (MacLaurin spheroid) becomes secularly unstable with respect to
triaxial $\ell=m=2$ perturbations (bar mode)
if $T/W > 0.1375$ (Jacobi/Dedekind bifurcation point
in the MacLaurin sequence). For compressible bodies (still in the Newtonian
regime), this threshold can be lowered to $0.1275$ (Bonazzola et al. 1996). 
The threshold of the CFS instability decreases with the mode number $m$
for polar modes:
for a Newtonian polytrope of abiabatic index $\gamma = 2$, one
has $T/W>0.079$ for a $m=3$ polar mode, 
$T/W>0.058$ for $m=4$ and $T/W>0.046$ 
for $m=5$ (Stergioulas \& Friedman 1998). 
Higher order modes are likely to be damped out by viscosity.
 For axial r-modes, the CFS instability is generic, i.e. it occurs at
any rotation rate (Andersson 1998). Madsen (1998) has however argued that the
r-mode instability is suppressed by the quark matter bulk 
viscosity in new born strange stars. 

In the relativistic case, a ratio $T/W$ can be defined according to 
Friedman et al. (1986) prescription. 
This quantity reduces to the usual
kinetic to gravitational energy ratio at the Newtonian limit, whereas its
physical interpretation in the relativistic case is not so clear 
(in particular the value of $T/W$ is coordinate dependent). However
it can be used to measure the importance of rotational effects. 

For the viscosity-driven instability, general relativistic terms 
increase the threshold on $T/W$: it becomes as high as $T/W>0.26$ for
a compactification parameter $M/R=0.2$, 
according to the post-Newtonian analysis of Shapiro \& Zane (1998). 
This stabilizing effect of general relativity onto the viscosity-driven
bar instability has also been found by Bonazzola et al. (1998a).

On the contrary, general relativity decreases the threshold of the CFS 
instability, down to $T/W>0.065$ for $m=2$ polar mode 
(this mode being always stable in the Newtonian regime for 
$\gamma=2$),
$T/W>0.046$ for $m=3$, $T/W>0.035$ for $m=4$, and
$T/W>0.029$ for $m=5$, 
for a $\gamma=2$ polytrope [Stergioulas \& Friedman 1998,
see also Yoshida \& Eriguchi (1997)]. 

\begin{figure}
\centerline{\includegraphics[width=8.5cm]{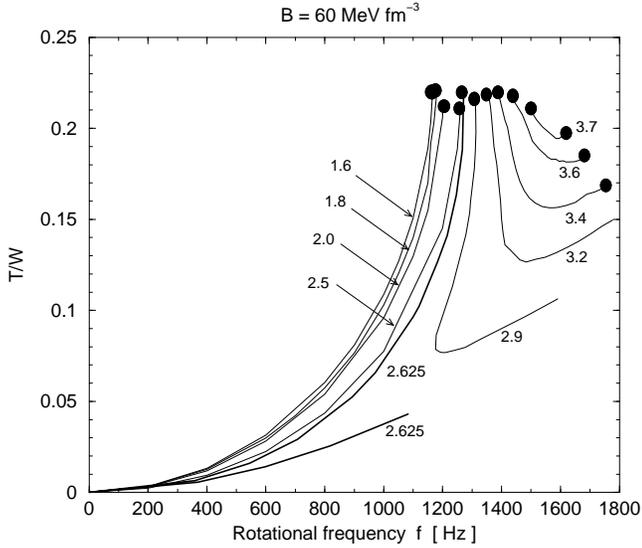}}
\caption[]{\label{f:T/W}
Kinetic to gravitational energy ratio $T/W$ as a function of the 
rotation frequency $f=\Omega/(2\pi)$ along sequences
of constant baryon number. Each sequence is labelled by its baryon mass in
${\rm M}_\odot$ units. The heavy black circles denotes Keplerian configurations.}
\end{figure}

\subsection{$T/W$ ratio of rotating strange stars}

Having the above results in mind, we have computed the $T/W$ ratio
according to the prescription of Friedman et al. (1986), 
for each of our rotating stange star models. 
In particular, Fig.~\ref{f:T/W} shows the value of $T/W$ as a function of
the rotation frequency along evolutionary sequences.
It appears clearly that the values of $T/W$ for strange stars are 
much greater
than for neutron stars (compare e.g. with Fig.~4 of Cook et al. (1994)). 
In particular, for the $\Omega_{\rm max}$ configuration, 
\begin{equation}
\left({T\over W}\right)_{\Omega_{\rm max}}
= 0.201~,
\end{equation}
(it does not depend on $B$) 
whereas it ranges from $0.105$ to $0.139$ for the set of neutron star EOS
examined in Nozawa et al. (1998). The fact that $T/W$ for maximally 
rotating strange stars is significantly larger than that for 
neutron stars, has been noted by Lattimer et al. (1990) and 
Colpi \& Miller (1992).  

Moreover, contrary to neutron stars, the value of $T/W$ remains very
high, of the order of $0.2$, for rapidly rotating strange stars 
of moderate mass: the value $T/W=0.2$ is reached at a rotation period
$P=0.86{\rm\  ms}$ (resp. $P=0.85{\ \rm ms}$) 
for a baryon mass of $M_{\rm B}=1.60\, {\rm M}_\odot$ 
(resp. $M_{\rm B}=1.80\, {\rm M}_\odot$), which
corresponds to a gravitational mass $M=1.31\, {\rm M}_\odot$ 
(resp. $M=1.47\, {\rm M}_\odot$) (notice that in the 
static case $M_{\rm B}=1.6,~1.8~{\rm M}_\odot$ for normal sequences in 
Fig.~\ref{f:T/W}  correspond to 
gravitational masses $M=1.27,~1.42~{\rm M}_\odot$, respectively). 

All this seems to indicate that triaxial instabilities could develop
more easily in rotating strange stars. 
However, it has to be stressed that
no actual stability analysis has been performed yet for strange stars which, 
in contrast to ordinary neutron stars, are bound not only 
by gravity, but also by confinement forces. These 
confinement forces are represented represented by the bag pressure, 
$-B$, acting on the strange star surface. 
The density profiles, Fig.~\ref{f:prof_dens1}, and adiabatic index $\gamma$, 
Fig.~\ref{f:gamma}, are qualitatively different from those characteristic of 
rotating neutron star models, for which triaxial instabilities have been 
studied. 
%
\section{Discussion and conclusions} \label{s:concl}
The numerical method used in the present paper is particularly suitable 
for exact calculations of models of rapidly rotating strange stars. 
In contrast to numerical approaches  used in previous calculations, the 
multi-domain  method enabled us to treat exactly the 
huge density discontinuity at the surface of strange stars. We used   
three-domain grid  in both $r$ and $\theta$ variables, 
adapted (adjusted) via a self-consistent procedure to the stationary 
shape of rotating strange star surface. The precision of our calculations 
was checked using several tests. We found significant differences between 
our results, and those obtained by  other authors 
 in previous exact calculations of the 
models of maximally rotating strange star models.

As noted in the early papers on the properties of strange stars,  
models  of static massive ($M\ga {\rm M}_\odot$)
 strange stars are characterized 
by global stellar parameters (mass, radius, moment of inertia, surface 
redshift) which are rather similar to those of ordinary neutron stars 
(Haensel et al. 1986, Alcock et al. 1986). 
In contrast to the static case,  
rapidly rotating strange stars exhibit qualitative differences with respect 
to rapidly rotating neutron stars.

 Strange stars rotating close to the 
break up (Keplerian)  frequency have kinetic to gravitational energy ratio 
$T/W\simeq 0.2$, which is nearly twice larger than corresponding values for 
neutron stars. Moreover, this unusually high 
  $T/W\sim 0.2$ is characteristic not exclusively  
for   supermassive rotating  models, but is typical also for  
normal rotating configurations 
 with baryon mass lower than the maximum allowed mass for 
static models. In particular, $T/W\simeq0.2$  is reached for 
   the $1.4~{\rm M}_\odot$ 
 models rotating at $P\simeq 0.85~$ms.    
Large value of $T/W$ stems from a  
 flat density profile  combined with  
strong equatorial flattening of rapidly rotating strange stars. 
These particular features of rapidly rotating strange stars are 
consequences of a special character of 
 EOS of strange quark matter in the strange 
star interior, reflected by a specific density dependence of the adiabatic 
index.  Strange  matter inside massive  ($M>{\rm M}_\odot$) strange stars 
  is very stiff in the outer (surface) 
layers and soft in the central core.  Large value of $T/W$ 
may  signal 
a relative softness (susceptibility) of rapidly rotating strange stars to 
triaxial instabilities. 

For simple  strange matter EOS, used in the present paper, 
several exact scaling relations hold for extremal (maximum frequency, 
maximum mass) configurations of rotating strange stars. 
Maximum rotation frequency as well as maximum mass  of rotating configurations 
exhibit an exact scaling with respect to the value of the bag constant. 
Empirical formula for $\Omega_{\rm max}$ for neutron stars, which relates  
this quantity to the mass and radius of static configuration with maximum 
allowable mass, holds with a high precision also for strange stars. 

Minimum period of rotating strange stars, stable with respect to 
axisymmetric perturbations, obtained for an acceptable range of the bag 
constant,  is $\simeq 0.5~$ms, and therefore similar to that characteristic 
of realistic models of neutron stars. However,  the effect of rapid rotation 
on the maximum allowable mass is significantly larger for strange stars 
than for neutron stars. Uniform rotation can increase the massimum allowable 
mass of strange stars by more than  $40\%$, to be compared with only 
 20\% increase for realistic models of neutron stars.

It should be stressed, that our numerical results for rapidly
rotating strange stars, summarized in Eqs.(24,25), and
Eqs.(27,28,29), have been obtained for a schematic EOS, Eq.(3).
This particular EOS has been selected for the sake of simplicity
and, last but not the least, for numerical convenience.  The most
general  - generic - feature of an EOS of strange matter is the
existence of a self-bound state at zero pressure, with
$E_0<930.4$~MeV. However, the actual EOS of strange matter -
should it exist - is expected to be different from that given by
Eq.(3). Therefore, results of the present paper neither represent
the full range of possibilities for ``theoretical'' strange stars,
nor they are  expected to correspond  to precise actual  values of
parameters of rapidly rotating strange stars - should they exist
in Nature. 

The equation of state of strange matter, used in the present paper,  
has been derived  neglecting  strange quark mass and representing 
 all effects
of QCD interactions  by  one parameter - the bag constant $B$.  
However, this simplest (minimal) model exhibits  generic properties 
of the EOS of a self-bound quark matter.  Generally, inclusion of 
perturbative corrections, expressed in terms of the QCD coupling 
constant $\alpha_{\rm c}$, and of the strange quark mass $m_{s}$, imply 
shifting down and simultaneous narrowing of the window of the values of 
$B$  compatible with strange matter hypothesis (Farhi \& Jaffe 1984). 
Typical values of $m_s$ are $\sim 200~{\rm MeV}/c^2$; the effect of 
$\alpha_{\rm c}\la 0.3$  on the EOS is then much smaller than that 
of $m_s$. Scaling relations  between extremal strange star configurations, 
corresponding to different values of $B$ 
(at fixed values of $\alpha_{\rm c}$, $m_s$) are than no longer exact. 
In the case of static strange star models, 
they are still very  precise (after the prefactors 
in these relations have been recalculated  to include the effect of 
$\alpha_{\rm c}$ and $m_s$) and therefore  useful (Haensel et al. 1986); 
similar situation is expected to take place in the case of rotating 
strange star models. 
In the static case, the effect of non-zero $m_s$ and $\alpha_{\rm c}$  
consisted in making the maximum mass configurations less massive but 
more compact, and  
 we expect similar effect for the  extremal 
configurations of rotating strange stars.

In this paper we restricted ourselves to the case of strange stars 
with superdense quark surface, built exclusively of strange quark matter
(bare strange stars). In principle, one should consider also models of 
strange stars, covered  by an envelope (crust) consisting of nuclei immersed 
in electron gas (Alcock et al. 1986). 
The nuclei, forming a crystal lattice of the crust, are separated  from  
strange matter by a repulsive coulomb barrier. 
 Both coulomb barrier, and characteristic spatial gap between nuclei and 
strange matter,   result from an electric dipole layer on the quark 
surface,  due to the nonuniform  density distribution of electrons near the 
quark surface. The presence of electrons in strange matter, 
necessary for equilibrium of quark core core with crust, results from 
nonzero strange quark mass. The maximum mass of the crust on a
 $M>{\rm M_\odot}$ strange star was originally estimated as  
$\sim 10^{-5}~{\rm M_\odot}$ (Alcock et al. 1986). Further studies of the 
crust--strange matter coexistence conditions led to an even lower 
value of the maximum mass of the crust on a $M>{\rm M}_\odot$ strange 
star, of the order of $10^{-6}~{\rm M}_\odot$  (Huang \& Lu 1997).  
 
In the case of the static strange star models, the effect of the presence 
of the crust on the maximum allowable mass configuration is very small. 
Rapid rotation will increase the maximum mass of a strange star crust 
(Glendenning \& Weber 1992), but still its effect on the maximally rotating 
and maximum mass configurations may be expected to 
be small:  both $\Omega_{\rm max}$ 
and $M_{\rm max}^{\rm rot}$ will be slightly reduced, as compared 
with the case of bare strange stars. It should be stressed, that the only 
existing calculation of rapid rotation 
of strange stars with crust  (Glendenning \& Weber 1992) was 
performed within a  version of   slow rotation approximation. 
The  multi-domain  method with adaptable grid, used in the present paper, 
is also   suitable for an exact treatment of rapid rotation of strange stars 
with crust, in which matter density exhibits a huge discontinuity at the
quark matter--crust interface, with outward density drop by more than three 
orders of magnitude. This problem is now being investigated.  

\begin{acknowledgements}
During his stay at DARC, Observatoire de Paris, P. Haensel was supported 
by the PAST professorship of French MENRT. 
This research was partially supported by the KBN grant No. 2P03D.014.13.
The numerical calculations have been performed on computers purchased
thanks to a special grant from the SPM and SDU departments of CNRS. 
\end{acknowledgements}

\end{document}